\documentclass[onecolumn,showpacs,preprintnumbers,amsmath,amssymb,superscriptaddress]{revtex4}

\usepackage{graphicx}
\usepackage{dcolumn,amssymb,amsmath,amsfonts}
\usepackage{bm}

\newcommand{\ket}[1]{\ensuremath{\left|#1\right\rangle}}
\newcommand{\bra}[1]{\ensuremath{\left\langle#1\right|}}

\newcommand{\be}{\begin{equation}}
\newcommand{\ee}{\end{equation}}
\newcommand{\bea}{\begin{eqnarray}}
\newcommand{\eea}{\end{eqnarray}}
\newcommand{\half}{\mbox{$\textstyle \frac{1}{2}$}}

\newcommand{\sdhalf}{\mbox{$\textstyle \sqrt{\frac{D}{2}}$}}
\newcommand{\sdthird}{\mbox{$\textstyle \sqrt{\frac{D}{3}}$}}
\newcommand{\sdd}{\mbox{$\textstyle \sqrt{\frac{D}{d-1}}$}}

\newcommand{\sca}[2]{\<#1 \ket{#2}}

\newcommand{\calu}{\mbox{$\cal U$}}
\def\a{\alpha}

\def\<{\langle}
\def\>{\rangle}

\begin{document}

\title{Robustness of a quantum key distribution with two
and three mutually unbiased bases}

\author{F. Caruso}
\affiliation{Scuola Superiore di Catania, via S. Paolo 73, I-95123 Catania, Italy.}

\author{H. Bechmann-Pasquinucci}

\affiliation{Dipartimento di Fisica "A. Volta", Unit\`a di
Pavia, Via Bassi 6, I-27100 Pavia, Italy.}
\affiliation{{\rm UCCI.IT}, via Olmo 26, I-23888 Rovagnate, Italy.}

\author{C. Macchiavello}
\affiliation{Dipartimento di Fisica "A. Volta", Unit\`a di
Pavia, Via Bassi 6, I-27100 Pavia, Italy.}

\date{\today}

\begin{abstract}
We study the robustness of various protocols for
quantum key distribution. We first consider the case of qutrits and study
quantum protocols that employ two and three mutually unbiased bases. We then
derive
the optimal eavesdropping strategy for two mutually unbiased bases in
dimension four and generalize the result to a quantum key distribution
protocol that uses two mutually unbiased bases in arbitrary finite dimension.
\end{abstract}


\maketitle

\section{\label{introduction}Introduction}
In the field of quantum information, quantum key distribution is
the application which is more developed, to the point that already
commercial prototypes exist. This fact is a good indicator of just
how much attention the subject has received in the last years. It
is therefore of primary importance to analyze in detail the
security of the various schemes proposed.
After the BB84 protocol, suggested originally by Bennett and Brassard in
1984 \cite{BB84} and based on the transmission of single qubits,
many generalized quantum key distribution
protocols appeared in the literature: the six-state protocol for
qubits
\cite{bruss,helle2}, the generalization to qutrits in \cite{helle}
and to ququarts in \cite{helle1}, and then subsequent generalizations
to arbitrary dimensions \cite{chiara,cerf}.
Several aspects of the security of these protocols have already been
analyzed \cite{fuchs,bruss,helle2,chiara,cerf}. Here we limit our
attention to quantum key distribution protocols based on the transmission of
single particles, and do not consider entanglement based schemes \cite{ekert}.

In this paper we analyze incoherent symmetric eavesdropping
attacks on some generalizations of the BB84 and the six-state
protocol, in which we use three- and four-dimensional systems and
vary the number of mutually unbiased bases used. For each protocol
we consider, our main purpose is  to derive the eavesdropping
strategy that is optimal with respect to the mutual information
shared between Alice and Eve, $I_{AE}$, for some given disturbance
$D$. This allows us to compare the robustness against
eavesdropping of the various protocols.

The paper is arranged in the following way: in the next section we
introduce the most general eavesdropping strategy for a set of
three-dimensional states and we impose the unitary and symmetry
conditions. In Subsec. \ref{2bases} we consider the cryptographic
protocol suggested in \cite{helle}, where, however, we use only
two rather than four mutually unbiased bases for coding the
information that Alice wants to communicate to Bob. In Subsec.
\ref{3bases} we analyze the same scheme but with three mutually
unbiased bases. In these sections we find the optimal
eavesdropping strategy and we compare the results with those ones
obtained by an optimal quantum cloning machine. The generalization
to four-dimensional systems and the comparison with the
corresponding results derived from cloning attacks \cite{cerf,cerf1,durt}
is discussed in Sec. \ref{ququart}, in
a protocol with only two unbiased bases. Finally, in the Sec.
\ref{qudit} we generalize the analysis to the case
of two mutually unbiased bases
with $d$-dimensional quantum states with arbitrary finite $d$.

\section{\label{qutrit}Optimal Eavesdropping with Three-Dimensional
Quantum States}

In this Section we derive the optimal incoherent eavesdropping strategies for
quantum cryptographic protocols based on the transmission of three-dimensional
systems (qutrits) and with two and three mutually unbiased bases.
We consider the general scenario where an eavesdropper
intercepts the quantum system in transit from Alice to Bob,
couples it to an ancilla by a
unitary interaction, and then forwards the original, but now
disturbed, quantum system to Bob while keeping the ancilla.
We assume that Eve can store the ancilla until the public discussion
between
Alice and Bob has taken place, since during their discussion the
measurement basis for each qutrit is
revealed.

The amount of information Eve can obtain from her system is
determined by the strength of the interaction, and how she later
measures the ancilla. The stronger the interaction the more
information Eve can extract from the ancilla but with the cost of
inducing a larger and larger disturbance on the system that Bob
finally receives. There is therefore a certain trade off between
the information she can gain and the disturbance that she
introduces on the system in transit from Alice to Bob. In the
following we optimize the information gain for a given value of
the disturbance.

In three dimensions it has been shown \cite{helle,chiara} that a
generalization of the
six-state protocol, which uses four mutually unbiased bases, i.e. 12
states,  is more robust against eavesdropping than the two dimensional
counter part.
Here we analyze incoherent and symmetric attacks
on three-dimensional quantum states in protocols which use two and
three mutually unbiased bases.  We remind that the word incoherent refer to
an attack where Eve interacts with one system in transit at a time.

Conventionally the first basis of the protocol corresponds to
the computational basis,
i.e. in a three-dimensional Hilbert space we denote it
by $\{\ket{0},\ket{1},\ket{2}\}$.
Then, the most general symmetric
eavesdropping strategy for qutrits is of the form
 \bea
\label{u} \ket{0}\ket{E}& \stackrel{\calu}{\longrightarrow} &
\sqrt{1-D}\ket{0}\ket{E_{00}}+\sdhalf
       \ket{1}\ket{E_{01}}+\sdhalf\ket{2}
           \ket{E_{02}},\nonumber  \\
\ket{1}\ket{E}&  \stackrel{\calu}{\longrightarrow} &
\sdhalf\ket{0}\ket{E_{10}}+\sqrt{1-D}\ket{1}
           \ket{E_{11}}+\sdhalf\ket{2}
           \ket{E_{12}},\\
\ket{2}\ket{E}&  \stackrel{\calu}{\longrightarrow} &
\sdhalf\ket{0}\ket{E_{20}}+\sdhalf\ket{1}
           \ket{E_{21}}+\sqrt{1-D}\ket{2}
           \ket{E_{22}},  \nonumber
\eea where $D$ is the disturbance introduced by Eve and $F=1-D$
represents the fidelity of the state that arrives at Bob
after the eavesdropping attack. We have indicated with $\ket{E}$
the initial state of Eve's system, while her states after the
interaction are denoted $\ket{E_{00}},\ket{E_{10}},\cdots$ and are
all normalized. We point out that the dimension of the Hilbert
space related to Eve's system is not fixed.
\newline
In order to satisfy the unitarity of $\calu$, the scalar products
between Eve's output states have to obey relations of the form \be
\sqrt{\frac{D(1-D)}{2}}(\sca{E_{ij}}{E_{jj}}+\sca{E_{ii}}{E_{ji}})+
\frac{D}{2}\sca{E_{ik}}{E_{jk}}=0,\nonumber  \ee where
$i=0,j=1,k=2$ and cyclic permutations.
 The requirement of symmetry reduces considerably the
complexity of the analysis, because it reduces the number of
parameters necessary to describe the most general eavesdropping
attack. Moreover, it has been shown \cite{gisin-cirac} that the
symmetry argument can be applied without lack of generalization.
The symmetry condition imposes some restrictions on the scalar
products which characterize the unitary operation $\calu$ used in
Eve's eavesdropping strategy: the scalar products between the
Eve's output states have to be invariant under the exchange of the
indices ($0$, $1$ and $2$) in order to treat the computational
basis states equally. Therefore it is possible to divide the
scalar products into 6 different groups, each group defining a
free parameter  ($x$, $y$, $z$, $t$, $w$ and $s$). In the
following the index is $i,j,k=0,1,2$:
\begin{itemize}
\item[$~$] $\sca{E_{ii}}{E_{ij}}=x$, for  $j\neq i$,
\item[$~$]$\sca{E_{ii}}{E_{jk}}=y$, where $i,j,k$ are all
different, \item[$~$] $\sca{E_{ij}}{E_{ik}}=z$, where $i,j,k$ are
all different, \item[$~$] $\sca{E_{ij}}{E_{ji}}=t$, for $j\neq i$,
\item[$~$] $\sca{E_{ij}}{E_{ki}}=w$, where $i,j,k$ are all
different, \item[$~$] $\sca{E_{ii}}{E_{jj}}=s$, for $j\neq i$; $s$
is a real number.
\end{itemize}
We will now specify the above strategy
for various protocols using different numbers of mutually unbiased
bases.

\subsection{Two mutually unbiased bases}\label{2bases}
\noindent
First we consider the cryptographic protocol suggested in
\cite{helle} with only two mutually unbiased
bases, namely a generalization of the BB84 protocol to dimension $d=3$.
We choose the second
basis to be the discrete
Fourier transform of the computational basis
 \bea
&&\ket{0'} =\frac{1}{\sqrt{3}}(\ket{0}+\ket{1}+\ket{2}), \nonumber \\
&&\ket{1'} =\frac{1}{\sqrt{3}}(\ket{0}+\a
\ket{1}+\a^*\ket{2}),  \\
&&\ket{2'} =\frac{1}{\sqrt{3}}(\ket{0}+\a^* \ket{1}+\a\ket{2}),
\nonumber \eea where $\a = e^{\frac {2\pi i}{3}}$. These two bases
are mutually unbiased since $|\sca{i}{j'}|=1/\sqrt{3}$ with
$i,j=0,1,2$.
\newline
We derive the optimal eavesdropping strategy for the quantum key
distribution protocol which uses these two bases by imposing the
same symmetry and unitarity conditions to the second basis of the
protocol as it was done for the first basis. This further reduces
the number of the parameters necessary to define the mutual
information between Alice and Eve.

The disturbance introduced by Eve to all possible input quantum
states has the following form
 \be \label{disturbance}
D_{(i)}=1-F_{(i)}= 1-\bra{i}\varrho^{(i)}_{B,out} \ket{i}, \ee
 \newline
where $\ket{i}$ is one of the possible states sent by Alice and
$\varrho^{(i)}_{B,out}=\mbox{Tr}_E
[\calu(\ket{i}\ket{E})(\bra{E}\bra{i})\calu^\dagger]$ is the
reduced density operator of the corresponding state sent on to Bob
after the interaction with Eve. By imposing that the disturbance
$D_{(i)}$ takes the same value $D$ for all 6 possible input
states, and by writing the disturbance introduced through the
eavesdropping transformation (\ref{u}) as a function of the scalar
products of Eve's output states, we find the following simple
relation among $w$, $D$ and $s$: \be \label{s} s=\frac{1-D
\mbox{Re}(w)}{1-D}-\frac{3D}{2(1-D)}. \ee For simplicity, we
consider $w$ to be real because only its real part appears in Eq.
(\ref{s}). Moreover, by imposing all the conditions discussed
above, we can conclude that the remaining four groups of scalar
products are zero, i.e. $x=y=z=t=0$ and $\sca{E_{ij}}{E_{jj}}=0$
with $i,j=0,1,2$ ($i \neq j$). We can now identify three
orthogonal sets of output states, $\{
\ket{E_{00}},\ket{E_{11}},\ket{E_{22}} \}$, $\{
\ket{E_{01}},\ket{E_{12}},\ket{E_{20}} \}$, $\{
\ket{E_{02}},\ket{E_{10}},\ket{E_{21}} \}$. The first set
corresponds to the case where the state has arrived correctly to
Bob, which happens with probability $F$. The second and the third
correspond to the cases where Bob obtains an error; in total this
happens with probability $D=1-F$. Notice, however, the difference
between the two sets of error states, the first of these sets
corresponds to Alice sending $i$ and Bob receiving $i+1$, whereas
the second corresponds to Alice sending $i$ and Bob receiving
$i+2$ {\it mod} 3.
 \newline
To describe the efficiency of an eavesdropping attack, we evaluate
the mutual information between Alice and Eve, which is the commonly used
figure of merit. We will derive
the optimal eavesdropping transformation for a fixed value $D$ of
the disturbance, by maximizing the mutual information $I_{AE}$
with respect to the free parameters of the strategy (i.e. the
non-trivial scalar products between Eve's output states). In
order to derive the expression of the mutual information
between Alice and Eve, we introduce the general parametrization for the
normalized output states
  \bea   \label{abc}
 \ket{E_{00}}&=&u\ket{\bar 0}+v\ket{\bar 1}+v\ket{\bar 2}, \nonumber \\
  \ket{E_{11}}&=&v\ket{\bar 0}+u\ket{\bar 1}+v\ket{\bar 2},  \\
  \ket{E_{22}}&=&v\ket{\bar 0}+v\ket{\bar 1}+u\ket{\bar 2}.\nonumber
 \eea
Since $s$ is real, in the above parametrization  we can take the
coefficients $u$ and $v$ to be real. Let us point out that in Eqs.
(\ref{abc}) $\{\ket{\bar 0},\ket{\bar 1},\ket{\bar 2}\}$
represents an orthonormal basis, orthogonal to all the other
output states of Eve's system, and that this particular
parametrization is due to the fact that, according to the symmetry
conditions imposed above, the overlaps of these three states must
be equal. Eve later uses a standard von Neumann measurement
\cite{singapore} on the basis $\{\ket{\bar i}\}$ to distinguish
these states. If the outcome of her measurement is the state
$\{\ket{\bar 0}\}$, she will interpret this as if the state was
$\ket{E_{00}}$, etc. In this way her probability of guessing the
state correctly is $u^2$, and the total probability for making an
error is $1-u^2=2v^2$.

Furthermore, we assume that the other two sets of states,
$\{\ket{E_{01}},\ket{E_{12}},\ket{E_{20}}\}$ and
$\{\ket{E_{02}},\ket{E_{10}},\ket{E_{21}}\}$, are parametrized in
a similar way where, instead of $u$ and $v$, we find,
respectively, two other real numbers, $r$ and $q$, and the basis
$\{\ket{\bar i}\}$ is replaced by two other orthogonal bases
$\{\hat{\ket{i}}\}$  and $\{\ket{\tilde i}\}$. Therefore Eve lets
her system interact with the state in transit according to Eqs.
(\ref{u}) and then, after listening to the public discussion
between Alice and Bob,  she performs a measurement. Eve's
probability of guessing the qutrit correctly when Bob received it
undisturbed is $u^2$, and when Bob's state is disturbed her
probability for guessing the qutrit correctly is $r^2$. This makes
it possible to compute Eve's probability of guessing the qutrit
correctly, $P(E)$,
\begin{eqnarray}
P(E)=F \ u^2 + D \ r^2. \label{eq:PE}
\end{eqnarray}

By using the symmetry conditions $2uv+v^2=s$ and $2rq+q^2=w$, and
by exploiting Eq. (\ref{s}), $u^2$ and $r^2$ can be expressed as
functions of $D$ and $w$
 \bea \label{fg}
& & u^2\equiv
\phi_3(D,w)=\frac{3+2D(w-1)}{9(1-D)}+
\frac{2\sqrt{2D[3-2D(2+w)](1+2w)}}{9(1-D)}, \nonumber \\
& & r^2\equiv
\lambda_3(w)=\frac{1}{9}\Big(5-2w+4\sqrt{1+w-2w^2}\Big).
  \eea

Based on these probabilities it is now possible to compute the
mutual information between Alice and Eve, $I_{AE,3}$ in terms of
the disturbance $D$ and the parameter $w$. There are two different
cases: (1) the qutrit has arrived correctly to Bob; this happens
with probability $F=1-D$, in which case Eve has probability
$\phi_3(D,w)$ for guessing the state correctly, (2) Bob has gotten
an error, this happens with probability $D$, in which case Eve has
probability $\lambda_3(w)$ for guessing the state correctly. This
means that the mutual information between Alice and Eve becomes
\begin{eqnarray}
I_{AE,3}=F \ I_3(\phi_3(D,w)) +D \ I_3(\lambda_3(w)),
\end{eqnarray}
where $I_3(x)=1+H_3(x)=1+x \log_3 x + (1-x) \log_3[(1-x)/2]$.
 Hence,
\bea
 I_{AE,3}(D,w)=1&+&(1-D)\Bigg[\phi_3(D,w)\log_3
 \phi_3(D,w)+[1-\phi_3(D,w)]
\log_3\frac{1-\phi_3(D,w)}{2}\Bigg]+\nonumber \\
 &+&D\Bigg[\lambda_3(w)\log_3 \lambda_3(w)+
[1-\lambda_3(w)]\log_3\frac{1-\lambda_3(w)}{2}\Bigg],
 \eea
where we have used the relation $F=1-D$. Through cumbersome calculations
(see the Appendix \ref{optimization} for details), we can prove
that, for fixed D, $I_{AE,3}(D,w)$  is maximized in correspondence
of the value $\bar w=\frac{3}{2}(\frac{2}{3}-D)$, which corresponds to
$\phi_3(D,\bar w)=\lambda_3(\bar w)$
and therefore
$I_{AE,3}(D, \bar w)$ is the optimal mutual information between
Alice and Eve and takes the following simple form: \bea
 I_{AE,3}(D,\bar w)=1+\phi_3(D,\bar w)\log_3 \phi_3(D,\bar w)+[1-\phi_3(D,\bar w)]\log_3\frac{1-\phi_3(D,\bar w)}{2}\;.
 \eea
Notice that Eve needs to employ an ancilla with dimension nine, or
equivalently two three-level systems, to implement the optimal attack.

As regards Bob, his mutual information with Alice decreases with
increasing disturbance as follows
 \be
 I_{AB,3}(D)=1+(1-D)\log_3(1-D)+D\log_3\frac{D}{2}\;.
\label{IAB}
 \ee

These results are plotted in Fig. \ref{fig:qutrit}. As we can see,
the information curves for Bob and Eve intersect at the critical
value for the disturbance $D_{c,2}= 0.2113$. For any value of
the disturbance smaller than this critical value the protocol is
guaranteed to be secure \cite{csiszar}.

\subsection{Three mutually unbiased bases}\label{3bases}

We now derive the optimal strategy for an extension of the above
protocol, namely with three rather than two mutually unbiased
bases. As before, the first basis is conventionally chosen as the
computational basis $\{\ket{0},\ket{1},\ket{2}\}$, while the
second basis is now defined as
 \bea \ket{0''}
&=&\frac{1}{\sqrt{3}}(\a
\ket{0}+\ket{1}+\ket{2}), \nonumber \\
\ket{1''} &=&\frac{1}{\sqrt{3}}(\ket{0}+\a
\ket{1}+\ket{2}), \nonumber \\
\ket{2''} &=&\frac{1}{\sqrt{3}}(\ket{0}+\ket{1} +\a\ket{2}), \eea
where $\a = e^{\frac {2\pi i}{3}}$.
 \newline
Similarly, the third basis is obtained by substituting in the above equations
$\a$
with $\a^{*}$.  Even if in general in dimension higher than two
different sets of
mutually unbiased bases are not unitarily equivalent,
we have checked that our results do not depend on the choice of the three
mutually unbiased bases, so we use these three
bases for convenience in the calculations.

Following the same procedure as above, we obtain the following simple
relation among $D$, $w$ and $s$
 \be s=\frac{1}{2}\frac{w D+2 -3D}{1-D}, \ee
 where $s$ and $w$ are defined at the beginning of Sec. \ref{qutrit}.
By imposing all the constraints as in the previous case, we can show
that $w$ is a real number and that all the other scalar products
are zero; in other words, there are again three orthogonal sets of states, $\{ \ket{E_{00}},\ket{E_{11}},\ket{E_{22}}
\}$, $\{ \ket{E_{01}},\ket{E_{12}},\ket{E_{20}} \}$, $\{
\ket{E_{02}},\ket{E_{10}},\ket{E_{21}} \}$ , i.e. $x=y=z=t=0$ and
$\sca{E_{ij}}{E_{jj}}=0$ with $i,j=1,2,3$ ($i \neq j$). \newline
In this case, we introduce the following general parametrization
for the set of normalized states of Eve
 \bea
 \ket{E_{00}} &=& u\ket{\bar 0}+v\ket{\bar 1}+v\ket{\bar 2}, \nonumber \\
  \ket{E_{11}} &=& v\ket{\bar 0}+u\ket{\bar 1}+v\ket{\bar 2}, \nonumber \\
  \ket{E_{22}} &=& v\ket{\bar 0}+v\ket{\bar 1}+u\ket{\bar 2},
  \eea
where, without loss of generality, we can take the coefficients to
be real. Again we assume that the other auxiliary states,
$\{\ket{E_{01}},\ket{E_{12}},\ket{E_{20}}\}$ and
$\{\ket{E_{02}},\ket{E_{10}},\ket{E_{21}}\}$, obey to the same
parametrization where, instead of $u$ and $v$, we find,
respectively, two other real numbers, $r$ and $q$.

Analogously to the previous case, after listening to the public
discussion between Alice and Eve,  Eve measures her system and the
probability of guessing the qutrit correctly, $P(E)$, is given by
Eq. (\ref{eq:PE}). By using the symmetry conditions, $2 u v +
v^2=s$ and $2 r q + q^2=w$, it is possible to express $u^2$ and
$r^2$ as functions of $D$ and $w$ as follows
 \bea
 u^2 \equiv \mu(D,w)&=&\frac{3-D(w+2)}{9(1-D)}+\frac{2\sqrt{2D[3+D(w-4)](1-w)}}{9(1-D)},\\
 r^2 \equiv \nu(w)&=&\frac{1}{9}\Big(5-2w+4\sqrt{1+w-2w^2}\Big). \nonumber
  \eea

Again, from these probabilities we can compute the mutual
information between Alice and Eve
\begin{eqnarray}
I_{AE,3}=F \ I_3(\mu(D,w)) +D \ I_3(\nu(w)).
\end{eqnarray}
The mutual information between Alice and Eve then takes
the explicit form
 \bea
 \tilde I_{AE,3}(D,w)=1&+&(1-D)\Bigg[\mu(D,w)\log_3 \mu(D,w)+
[1-\mu(D,w)]\log_3\frac{1-\mu(D,w)}{2}\Bigg]+\nonumber \\
 &+&D\Bigg[\nu(w)\log_3 \nu(w)+[1-\nu(w)]\log_3\frac{1-\nu(w)}{2}\Bigg].
 \eea

\begin{figure}[th]
\includegraphics[width=0.63\textwidth]{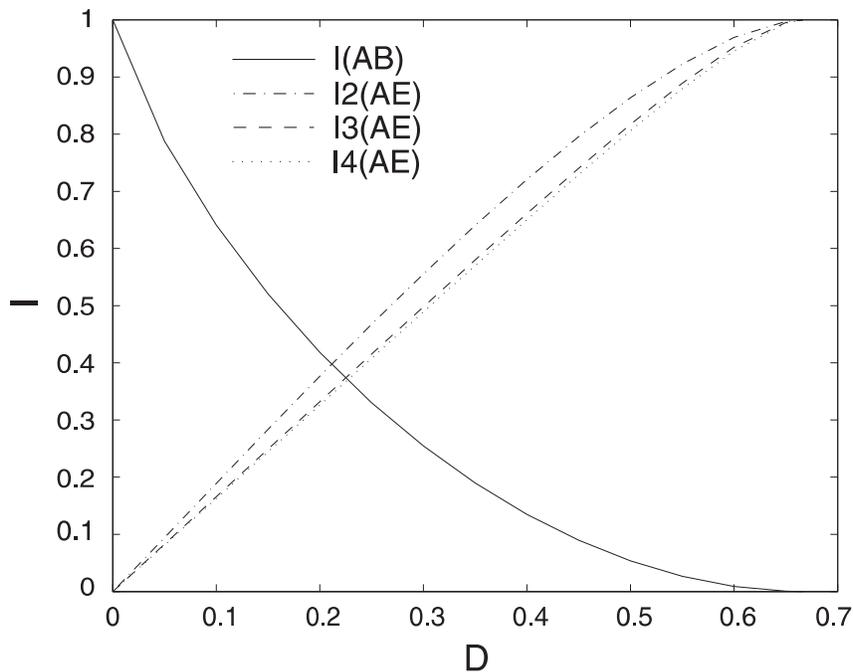}
\caption{\label{fig:qutrit} Mutual Information for Alice/Bob
(I(AB)) and Alice/Eve as a function of the disturbance $D$, for
three-dimensional quantum states in a scheme with two (I2(AE)),
three (I3(AE)) and four (I4(AE)) mutually unbiased bases. The
latter curve was derived in \cite{chiara}.}
\end{figure}

Notice that, in contrast with the previous case, here we have
found no simple analytical solution for the optimal mutual
information between Alice and Eve. Therefore in Fig.
\ref{fig:qutrit} we plot a numerical solution for the optimal
expression. The mutual information between Alice and Bob takes the
form (\ref{IAB}), as in the previous case. The information curves
for Bob and Eve intersect at a value of the disturbance
$D_{c,3}\simeq 0.2247$, which is larger than $D_{c,2}= 0.2113$.
These two values have also to be compared to the critical
value $D_{c,4}=0.2267$ \cite{chiara} of the protocol that
employs four mutually unbiased bases, which is the maximum number
in dimension three. \par  As expected, the critical value
increases for increasing number of mutually unbiased bases, but it
increases weakly. On the other hand the key generation rate
decreases (the key generation rate decreases as the inverse of the
number of bases employed). Therefore, in a realistic scenario the
optimal choice for the number of bases to be employed in a
protocol will depend on a convenient balance between the two
trends.

Finally, we compare the above results  with the ones obtained by
optimal quantum cloning attacks. The cases with two and four bases
are studied analytically in \cite{cerf}, while in
\cite{cerf1,durt} the case with three mutually unbiased bases is
studied numerically. The results obtained both with the most
general unitary eavesdropping strategy and with the optimal
quantum cloning machine are exactly the same. Therefore the
quantum cloning machine would represent an optimal eavesdropping
strategy for these quantum key distribution protocols.

\newpage
\section{\label{ququart}Optimal Eavesdropping with Four-Dimensional Quantum States}
We will now derive the optimal eavesdropping strategy for two
mutually unbiased bases in dimension $d=4$.
Now let us introduce the computational basis
$\{\ket{0},\ket{1},\ket{2},\ket{3}\}$
and write the most general unitary symmetric
eavesdropping strategy for a set of four-dimensional quantum
states (ququarts): \bea \label{uu}
  \ket{i}\ket{E} \stackrel{\calu}{\longrightarrow}
\sqrt{1-D}\ket{i}\ket{E_{ii}}+\sdthird
       \ket{i+1}\ket{E_{i \ i+1}}+\sdthird\ket{i+2}
         \ket{E_{i \ i+2}}+\sdthird\ket{i+3}
         \ket{E_{i \ i+3}},
\eea where $i=0,1,2,3$ and the index additions are taken modulo
$4$.

In order to satisfy the unitarity of $\calu$, the scalar products
between Eve's output states have to obey to constraints similar to
the three-dimensional case and, for the symmetry of the problem,
we have again a classification of Eve's output states into six
sets of scalar products, each defining a free parameter. The
scalar products have to fulfill the following conditions
\begin{itemize}
\item[$~$] $\sca{E_{ii}}{E_{ij}}=x$, for $i\neq j$, \item[$~$]
$\sca{E_{ii}}{E_{jk}}=y$, where $i,j,k$ are all different,
\item[$~$] $\sca{E_{ij}}{E_{ik}}=z$, where $i,j,k$ are all
different, \item[$~$] $\sca{E_{ij}}{E_{jh}}=t$, for $i,j,h$ all
different,
 \item[$~$] $\sca{E_{ij}}{E_{hk}}=w$, where
$j\neq i$, \ ($h = j$ and $k=i$) or ($h,k,i,j$ all different); it
turns out that $w$ is a real number,
 \item[$~$] $\sca{E_{ii}}{E_{jj}}=s$, for $i\neq j$; $s$ is also real.
\end{itemize}

Let us consider the protocol, suggested in \cite{helle1}, where
the first basis is the computational basis and the second basis,
connected by a discrete Fourier transform to the first one, is
defined as
 \bea \ket{0'} &=&\half(\ket{0}+\ket{1}+\ket{2}+\ket{3}), \nonumber \\
\ket{1'} &=&\half(\ket{0}-\ket{1}+\ket{2}-\ket{3}), \\
\ket{2'} &=&\half(\ket{0}-\ket{1}-\ket{2}+\ket{3}), \nonumber \\
\ket{3'} &=&\half(\ket{0}+\ket{1}-\ket{2}-\ket{3}). \nonumber
 \eea

By imposing the additional  conditions that the disturbance must
be the same for all eight possible states sent by Alice  the
number of free parameters is further reduced because it turns out
that $x=y=z=t=0$. We can then derive the following expression for
$s$ as a function of $D$ and $w$ \be s=\frac{1-w D}{1-D} +
\frac{4}{3} \frac{D}{D-1}\;. \label{s4}
 \ee
Again, we introduce the following parametrization for Eve's output states,
by generalizing the procedure followed in the three-dimensional case
  \bea
 \ket{E_{00}}&=&u\ket{\bar 0}+v\ket{\bar 1}+v\ket{\bar 2}+v\ket{\bar
3},\nonumber\\
  \ket{E_{11}}&=&v\ket{\bar 0}+u\ket{\bar 1}+v\ket{\bar 2}+v\ket{\bar 3},\\
 \ket{E_{22}}&=&v\ket{\bar 0}+v\ket{\bar 1}+u\ket{\bar 2}+v\ket{\bar
3},\nonumber\\
  \ket{E_{33}}&=&v\ket{\bar 0}+v\ket{\bar 1}+v\ket{\bar 2}+u\ket{\bar
3}\nonumber.
  \label{abcd}
 \eea
where $u$ and $v$ are real numbers.

Analogously to the three-dimensional case, a similar
parametrization (with $r$, $q$ real) is chosen for the other three
sets of states $\{\ket{E_{01}}, \ket{E_{10}}, \ket{E_{23}},
\ket{E_{32}}\}$, $\{\ket{E_{02}}, \ket{E_{13}}, \ket{E_{20}},
\ket{E_{31}} \}$ and $\{\ket{E_{03}}, \ket{E_{12}}, \ket{E_{21}},
\ket{E_{30}}\}$. After the public discussion between Alice and
Bob, Eve performs a measurement and has the following probability,
$P(E)$, to make the correct estimation of the qutrit
 \bea
P(E)=F \ u^2 + D \ r^2 ,
 \eea
where
 \bea
 u^2 \equiv \phi_4(D,w)&=&\frac{4-2D(1-3w)}{16(1-D)}+2\frac{\sqrt{3D(1+3w)
[4-D(5+3w)]}}{16(1-D)}, \\
 r^2 \equiv \lambda_4(w)&=&\frac{1}{8}\Big(5-3w+3\sqrt{1+2w-3{w}^2}\Big).
  \eea
The above expressions are calculated by exploiting the
normalization conditions, the symmetry conditions and using Eq.
(\ref{s4}). \par From the above probabilities the mutual
information between Alice and Eve can be derived
\begin{eqnarray}
I_{AE,4}=F \ I_4(\phi_4(D,w)) +D \ I_4(\lambda_4(w)),
\end{eqnarray}
where $I_4(x)=1+H_4(x)=1+x \log_4 x + (1-x) \log_4[(1-x)/3]$.
\par
The mutual information has now to be optimized as a function of $w$.
This can be done analytically, following the same procedure as in the three
dimensional case reported explicitly in the Appendix. It turns out that
the solution of the optimization is given by the value
$\bar w=\frac{4}{3}(\frac{3}{4}-D)$ and
the optimal mutual information between Alice and Eve is then given by
 \bea
 I_{AE,4}(D,\bar w)=1+\phi_4(D,\bar w)\log_4 \phi_4(D,\bar w)+[1-\phi_4
(D,\bar w)]\log_4\frac{1-\phi_4(D,\bar w)}{3}\;. \eea This
expression has to be compared with the mutual information between
Alice and Bob, which is now given by
 \be
 I_{AB,4}(D)=1+(1-D)\log_4(1-D)+D\log_4\frac{D}{3}\;.
 \ee
These curves are plotted in Fig. 2, and compared to the optimal
one corresponding to five mutually unbiased bases, conjectured in
\cite{chiara}. The two critical values are $D_{c,2}= 0.25$ for
two mutually unbiased bases and $D_{c,5}=0.2666$ for five
mutually unbiased bases. As we can see, the robustness of the
protocol increases as the number of mutually unbiased bases
increases, at the expense of a considerable reduction in the key
generation rate, which goes from 1/2 to 1/5.

Finally, we compare the results above with the ones corresponding
to the relative optimal quantum cloning machines, as studied in
\cite{cerf}; as in the case of qutrits, the curves for the optimal mutual
information between Alice and Eve  are the same.

\begin{figure}[th]
\includegraphics[width=0.615\textwidth]{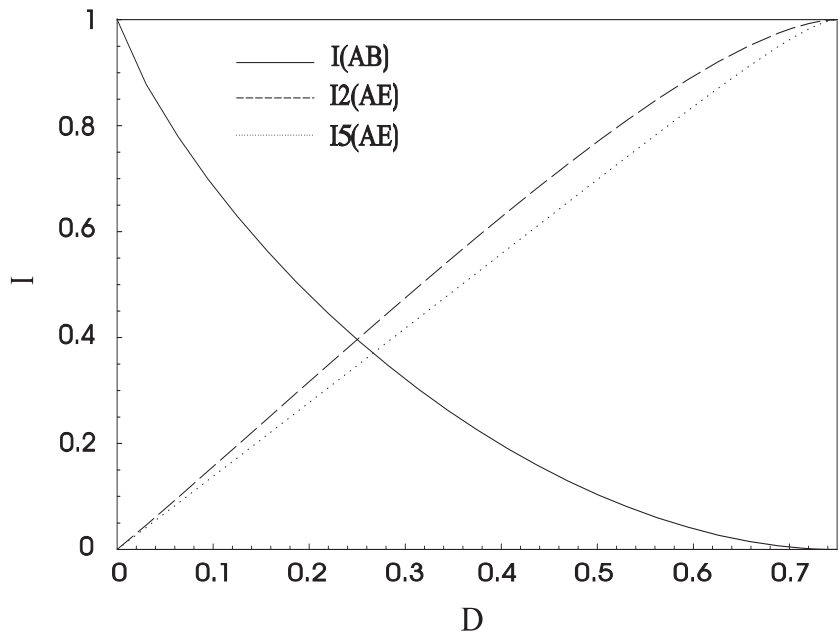}
\caption{\label{fig:ququart} Mutual Information for Alice/Bob
(I(AB)) and Alice/Eve as a function of the disturbance, for
four-dimensional quantum states in a scheme with two (I2(AE)) and
five (I5(AE)) mutually unbiased bases (the latter curve was
conjectured in \cite{chiara}) .}
\end{figure}

\newpage
\section{\label{qudit}Generalization to $d$-Dimensional Quantum States}

In this Section we generalize the above analysis to the case of
two mutually unbiased bases with $d$-dimensional quantum states
(qudits) with arbitrary finite $d$.
After introducing the computational basis $\{\ket{k}\}$ with
$k=0,...,d-1$, the most general symmetric eavesdropping strategy
for qudits takes the form \bea \label{uud}
   \ket{i}\ket{E} \stackrel{\calu}{\longrightarrow}  \sqrt{1-D}\ket{i}\ket{E_{ii}}+\sdd
 \ket{i+1}\ket{E_{i \ i+1}}+\sdd\ket{i+2}
         \ket{E_{i \ i+2}}+ \ ... \ + \sdd\ket{i+d-1}
         \ket{E_{i \ i+d-1}},
\eea where $i=0,1,2,3,...,d-1$ and the index additions are taken
modulo $d$. \par Now we consider the cryptographic protocol where
the two mutually unbiased bases are given by the computational
basis $\{\ket{0},\ket{1},...\}$ and its Fourier transformed
 \be
 \ket{\bar l}= \frac{1}{\sqrt{d}} \sum_{k=0}^{d-1}{e^{2 \pi i (k l/d)}
 \ket{k}},
  \ee with $l=0,...,d-1$.

We follow the same procedure
as in the previous sections, by imposing the symmetry conditions and
the requirements that all possible $2d$ states sent by Alice are
equally disturbed. In this way we obtained that many scalar
products among Eve's output states
are zero and the number of free parameters is reduced.
\par In particular, it is possible to divide Eve's output states
into $d$ orthogonal sets: one of these sets is
$\{\ket{E_{00}},\ket{E_{11}},...,\ket{E_{d-1 \ d-1}}\}$, while the
other ones assume a particular form according to the parity of the
dimension of the Hilbert space, $d$.
\par If $d$ is odd, the other $d-1$ sets are formed by
$\{\ket{E_{i1}},\ket{E_{i+1 \ 2}},\ket{E_{i+2 \
3}},...,\ket{E_{i+d-1 \ 0}}\}$, where $i=0,1,...,d-1$. Instead, if
the dimension $d$ of the Hilbert space is even, we have the
following $d-1$ sets:
\begin{itemize}

\item[$~$] $\{\ket{E_{0 j}},\ket{E_{1 \ j+1}},...,\ket{E_{d-1 \
j+d-1}}\}$ where $j=2,4,...,d-2$.

\item[] \item[$~$] $\{\ket{E_{0 j}},\ket{E_{1 \
j-1}},...,\ket{E_{d-1 \ j-d+1}}\}$ where $j=1,3,...,d-1$.

\end{itemize}
Independently of the parity of $d$, the scalar products between
any two states, belonging to one of these $d-1$ sets, are always
the same and equal to a free parameter $w$. Moreover, analogously to the case
of qutrits and ququarts, there is the
additional condition $\bra{E_{ii}}E_{jj}\rangle=s$ for $i\neq j$.

Combining all the constraints of the problem, the generalized
relation among $D$, $w$ and $s$ is as follows
 \be s=\frac{1-w D}{1-D} + \frac{d}{d-1} \frac{D}{D-1}\;.
 \ee
After introducing the proper parametrization for  Eve's output
states and the relative set of probabilities for her measurement,
we obtain the mutual information between Alice and Eve and then we
optimize it with respect to the free parameter $w$. Finally the
optimal mutual information between Alice and Eve has the following form
 \bea
 I_{AE,d}(D,\bar w)=1+\phi_d(D,\bar w)\log_d \phi_d(D,\bar w)+[1-
\phi_d(D,\bar w)]\log_d\frac{1-\phi_d(D,\bar w)}{d-1},
 \eea
 where
 \bea \phi_d(D,\bar w)=\frac{1}{d^2(1-D)}
 \Big[d+D[-2+(d-2)(d-1)\bar w]+2\sqrt{(d-1)D[1+(d-1)\bar w]\{d-D[1+d+(d-1)\bar w]\}}\Big], \eea
and \be \bar w=\frac{d}{d-1}\Bigg(\frac{d-1}{d}-D\Bigg). \nonumber
\ee
\par
For any value of $d$, the function $I_{AE,d}(D)$ is the same as
that one obtained in \cite{cerf} with a cloning-based attack.
This expression has to be compared with the mutual information
between Alice and Bob, which is given by
 \be
 I_{AB,d}(D)=1+(1-D)\log_d(1-D)+D\log_d\frac{D}{d-1}\;.
 \ee
Analogously to the cloning attack,
we find the following analytical expression for the
critical disturbance, $D_c$, as function of $d$
 \be
D_c(d)=\frac{1}{2}\Bigg(1-\frac{1}{\sqrt{d}} \Bigg).
 \ee
The above expression proves that the robustness of the quantum channel
increases as
the dimension of the quantum system used in the protocol
increases.

\section{\label{conclusions}Concluding Remarks}
In this paper we study some different quantum key distribution
protocols in order to compare their robustness against eavesdropping. A
protocol is said to be more robust if it tolerates a higher disturbance
and still allows Alice and Bob to generate a secure key.
We have derived the
optimal eavesdropping strategy for each protocol,
concentrating on symmetric and incoherent attacks and
evaluated the
optimal mutual information between Alice and Eve for a given disturbance.
We note that the
robustness of quantum key distribution increases with the dimension of the
space, hence reflecting the fact that since there are more states, an
error can be distributed among more states. Increasing the number of
mutually unbiased bases used also improves the robustness
against eavesdropping. However, increasing the number of bases has to be
weighted against a lower key generation rate since the probability that
Alice and Bob used the same basis goes down.

We have compared our results with the ones obtained under the
assumption that optimal
asymmetric quantum cloning machines \cite{cerf,cerf1,durt} provides
optimal eavesdropping. In this comparison we note that the values
of the critical disturbance are always the same and therefore we
can conclude that the quantum cloning machine represents the
optimal eavesdropping strategy of the quantum key distribution
protocols studied in this paper.

\appendix

\section{The optimal mutual information
Alice/Eve}\label{optimization}

In this Appendix we show the analytical calculations in order to
maximize the mutual information between Alice and Eve in a
protocol with two mutually unbiased bases. We analyze in detail
the three-dimensional case, but the proof can be easily extended
to higher dimensions.
 \par
Recall that in a protocol with two bases the mutual information
Alice/Eve for three-dimensional quantum states, as a function of the
disturbance ($D$) and the free parameter ($w$), has the following
analytical expression:
 \bea
 I_{AE,3}(D,w)=1&+&(1-D)\Bigg[\phi_3(D,w)\log_3
 \phi_3(D,w)+[1-\phi_3(D,w)]
\log_3\frac{1-\phi_3(D,w)}{2}\Bigg]+\nonumber \\
 &+&D\Bigg[\lambda_3(w)\log_3 \lambda_3(w)
+[1-\lambda_3(w)]\log_3\frac{1-\lambda_3(w)}{2}\Bigg],
 \eea
 where $\phi_3(D,w)$ and $\lambda_3(w)$ are given in Eqs. (\ref{fg}).
Using the normalization of the ancilla states and taking into
account the expression (\ref{s}), we can prove that
 \begin{eqnarray}
\partial_{w}I_{AE,3}(D,w) \Big|_{\bar w} = 0 \Longleftrightarrow \bar
 w=\frac{3}{2}\Bigg(\frac{2}{3}-D\Bigg).
\end{eqnarray}
\par
In fact, when $\bar w=\frac{3}{2}\Big(\frac{2}{3}-D\Big)$, the
following relations are satisfied
\begin{displaymath}
\left\{ \begin{array}{l}
\phi_3(D,\bar w) \equiv \lambda_3(\bar w)\\
\frac{\partial_{w}\phi_3(D,w)}{\partial_{w}\lambda_3(w)}\Big|_{\bar
  w}=\frac{D}{D-1}
\end{array} \Longrightarrow \partial_{w}I_{AE,3}(D,w) \Big|_{\bar w} = 0. \right .
\end{displaymath}
\par
Therefore the stationary points of $I_{AE,3}$ are on the plane
$\bar w=\frac{3}{2}\Big(\frac{2}{3}-D\Big)$ and, because of the
concavity of the mutual information Alice/Eve, $I_{AE,3}(D,\bar
w)$ is the maximal mutual information that Eve can extract from
the quantum channel Alice/Bob. The mutual information therefore has the
following expression:
\bea
 I_{AE,3}(D,\bar w)=1+\phi_3(D,\bar w)\log_3 \phi_3(D,\bar w)
+[1-\phi_3(D,\bar w)]\log_3\frac{1-\phi_3(D,\bar w)}{2},
 \eea
where $\bar w=\frac{3}{2}\Big(\frac{2}{3}-D\Big)$.

In order to prove the concavity of the mutual information
Alice/Eve, $I_{AE,3}(D,w)$, let us consider the auxiliary
two-variable function, $f(a,b)$, as follows
 \bea
f(a,b)=1+(1-D)\{a \log[a]+(1-a)\log[(1-a)/2
]\}+D\{b\log[b]+(1-b)\log[(1-b)/2 ]\}.
 \eea
Now let $a(D,w)\equiv \phi_3(D,w)$ and $b(w) \equiv \lambda_3(w)$.
Because the values of $\phi_3(D,w)$ and $\lambda_3(w)$ are in the
range $[\frac{1}{3},1]$, we have
 \be
 \partial_a{f(a,b)}=(1-D)\log[2a/(1-a)]>0 , \ \ \ \ \ \ \
 \partial_b{f(a,b)}=(1-D)\log[2b/(1-b)]>0 .
 \ee
Then the second derivatives are
 \bea
 \partial_{a,a}{f(a,b)}=\frac{1-D}{a(1-a)} > 0 , \ \ \ \ \ \ \ \ \
\partial_{b,b}{f(a,b)}=\frac{1-D}{b(1-b)} > 0 ,
 \eea
whence
 \bea
 \partial_{w}I_{AE,3}(D,w) \equiv \partial_{w}f(a,b)=\partial_a
 f(a,b)
 \partial_w a+ \partial_b f(a,b)
 \frac{d b}{d w}
 \eea
and
 \bea
 \partial_{w,w}I_{AE,3}(D,w) \equiv \partial_{w,w} f(a,b)=
 \partial_{a,a} f(a,b)
 \Big(\partial_w a\Big)^2+ \partial_a f(a,b) \partial_{w,w} a
+\partial_{b,b} f(a,b) \Big(\frac{d b}{d w}\Big)^2+
\partial_b f(a,b) \frac{d^2
 b}{d w^2}.
 \eea
Combining all these equations together, we obtain
 \bea
 \partial_{w,w}I_{AE,3}(D,w)=(1-D)\Bigg[\frac{1}{\phi(D,w) [1-\phi(D,w)]}
 [\partial_w{\phi(D,w)}]^2
+\log
\Bigg[\frac{2\phi(D,w)}{1-\phi(D,w)}\Bigg]\partial_{w,w}\phi(D,w)
\Bigg]+\nonumber \\
 +D \Bigg[\frac{1}{\lambda(w)[1-\lambda(w)]} [\lambda^{'}(w)]^2+
 \log \Bigg[\frac{2\lambda(w)}{1-\lambda(w)}\Bigg]\lambda^{''}(w) \Bigg] <
 0.
 \label{sec-der} \eea
Recall that $0 < D < 2/3$. The parentheses on the right hand side
of Eq. (\ref{sec-der}) are always negative in the domain of
definition of the function $I_{AE,3}(D,w)$. Therefore $
\partial_{w,w}I_{AE,3}(D,w)<0$ and it follows that the function $I_{AE,3}(D,w)$ is concave.

The generalization to arbitrary dimension, which leads to the solution
$\bar w=\frac{d}{d-1}(\frac{d-1}{d}-D)$, is then straightforward.

\begin{acknowledgments}
We wish to thank D. Bruss and N. Gisin for useful discussions.

This work has been supported in part by the EC under the project SECOQC
(contract n. IST-2003-506813).
\end{acknowledgments}

\end{document}